\documentclass[twoside]{ilcws07}
\usepackage[latin1]{inputenc}
\usepackage[dvips]{graphicx,epsfig,color}
\usepackage{wrapfig,rotating}
\usepackage{cite}
\usepackage{amssymb,amsmath,array}

\begin{document}
\title{First results on $e^+e^- \to 3$~jets at NNLO} 
\author{A.\ Gehrmann-De Ridder$^1$, T.\ Gehrmann$^2$, E.W.N.\ Glover$^3$,
G.\ Heinrich$^4$
\vspace{.3cm}\\
1- Institute for Theoretical Physics, ETH, CH-8093 Z\"urich,
Switzerland
\vspace{.1cm}\\
2- Institut f\"ur Theoretische Physik,
Universit\"at Z\"urich, CH-8057 Z\"urich, Switzerland
\vspace{.1cm}\\
3-  Institute of Particle Physics Phenomenology, 
        Department of Physics,\\
        University of Durham, Durham, DH1 3LE, UK
\vspace{.1cm}\\
4-  School of Physics, The University of Edinburgh, Edinburgh EH9 3JZ,
UK\\
}

\maketitle

\begin{abstract}
Precision studies of QCD at $e^+e^-$ colliders are based on measurements of 
event shapes and jet rates. To match the high experimental accuracy, 
theoretical predictions to next-to-next-to-leading order (NNLO) in QCD 
are needed for a reliable interpretation of the data. We report the
first calculation of NNLO corrections (${\cal O}(\alpha_s^3)$) to three-jet
production and related event shapes, and discuss their 
phenomenological impact. 
\end{abstract}

\vspace{-8cm}
{\noindent ZU-TH 21/07, IPPP/07/53, Edinburgh 2007/20}
\vspace{8.4cm}
\section{Introduction}
Measurements at LEP and at earlier $e^+e^-$ colliders have 
helped to establish QCD as the theory of strong interactions by 
directly observing gluon radiation through three-jet production events. 
The LEP measurements of three-jet production and related event shape 
observables are of a very high
statistical precision. The extraction of $\alpha_s$ from these
data sets relies on a comparison of the data with theoretical predictions.
Comparing the different sources of error in this extraction,
one finds that the experimental error is negligible compared to
the theoretical uncertainty. There are two sources of theoretical
uncertainty: the theoretical description of the parton-to-hadron
transition (hadronisation uncertainty) and the uncertainty stemming from the 
truncation of the perturbative series at a certain order, as estimated by scale
variations (perturbative or scale uncertainty). 
 Although the precise
size of the hadronisation uncertainty is debatable and perhaps often
underestimated, it is certainly appropriate to consider the scale
uncertainty as the dominant source of theoretical error on the precise
determination of  $\alpha_s$ from three-jet observables. From the 
planned luminosity of the ILC, one would expect measurements of event 
shapes comparable in statistical quality to what was obtained at LEP, 
thus allowing for precision QCD studies at ILC energies.

So far the three-jet rate and related event shapes
have been calculated~\cite{ert,ert2} up to the next-to-leading order (NLO),
improved by a resummation of leading and subleading infrared 
logarithms~\cite{ctwt,ctw} and by the inclusion of power
corrections~\cite{power}. 

QCD studies of event shape 
observables at LEP~\cite{expreview} are based around the use of 
NLO parton-level event generator 
programs~\cite{event}.  
As expected, the current error on $\alpha_s$ from these observables  
\cite{Bethke} is dominated by the theoretical uncertainty.
Clearly, to improve the determination of $\alpha_{s}$, 
the calculation of the NNLO corrections to these
observables becomes mandatory. We present here the first NNLO calculation 
of three-jet production and related event shape variables.

\section{Calculation}
Three-jet production at tree-level is induced by the decay of a virtual
photon (or other neutral gauge boson) into a quark-antiquark-gluon final
state. At higher orders, this process receives corrections from extra
real or virtual particles. The individual partonic channels
that contribute through to NNLO
are shown in Table~\ref{table:partons}. All of the tree-level and loop
amplitudes associated with these channels are known 
in the literature~\cite{3jme,muw2,V4p,tree5p}.

For a given partonic final state, jets are reconstructed according to 
the same definition as in the experiment, which is applied to partons instead 
of hadrons. At leading order, all three final state partons must be 
well separated from each other. At NLO, up to four partons 
can be present in the final state, 
two of which 
can be clustered together,   
whereas at NNLO, the final state can consist of up to five partons, 
such that as many as three partons can be clustered together. 
The more partons in the final state, 
the better one expects the matching between theory and 
experiment to be.

\begin{wraptable}{l}{0.5\columnwidth}
\centerline{
\begin{tabular}{lll}
\hline\\
LO & $\gamma^*\to q\,\bar qg$ & tree level \\[2mm]
NLO & $\gamma^*\to q\,\bar qg$ & one loop \\
 & $\gamma^*\to q\,\bar q\, gg$ & tree level \\
 & $\gamma^*\to q\,\bar q\, q\bar q$ & tree level \\[2mm]
NNLO & $\gamma^*\to q\,\bar qg$ & two loop \\
 & $\gamma^*\to q\,\bar q\, gg$ & one loop \\
& $\gamma^*\to q\,\bar q\, q\,\bar q$ & one loop \\
& $\gamma^*\to q\,\bar q\, q\,\bar q\, g$ & tree level \\
& $\gamma^*\to q\,\bar q\, g\,g\,g$ & tree level\\[2mm]
\hline
\end{tabular}
}
\caption{Partonic contributions to three-jet final states  in 
perturbative QCD.}
\label{table:partons}
\end{wraptable}

The two-loop $\gamma^* \to q\bar q g$ matrix elements were derived 
in~\cite{3jme} by reducing all relevant Feynman integrals to a small 
set of master integrals using integration-by-parts~\cite{ibp} and 
Lorentz invariance~\cite{gr} identities, solved with the Laporta 
algorithm~\cite{laporta}. The master integrals~\cite{3jmaster} were 
computed from their differential equations~\cite{gr} and expressed 
analytically
in terms of one- and two-dimensional harmonic polylogarithms~\cite{hpl}. 

The one-loop four-parton matrix elements relevant here~\cite{V4p} were 
originally derived in the context of NLO corrections to four-jet 
production and related event shapes~\cite{fourjetprog,cullen}. One of 
these four-jet parton-level event 
generator programs~\cite{cullen} is the starting point 
for our calculation, since it already contains all relevant 
four-parton and five-parton matrix elements.

The four-parton and five-parton contributions to three-jet-like final 
states at NNLO contain infrared real radiation singularities, which 
have to be extracted and combined with the 
infrared singularities~\cite{catani} 
present in the virtual three-parton and four-parton contributions to 
yield a finite result. In our case, this is accomplished by 
introducing subtraction functions, which account for the 
infrared real radiation singularities, and are sufficiently simple to 
be integrated analytically. Schematically, this subtraction reads:
\begin{eqnarray*}
\lefteqn{{\rm d}\sigma_{NNLO}=\int_{{\rm d}\Phi_{5}}\left({\rm d}\sigma^{R}_{NNLO}
-{\rm d}\sigma^{S}_{NNLO}\right) }\nonumber \\ 
&&+\int_{{\rm d}\Phi_{4}}\left({\rm d}\sigma^{V,1}_{NNLO}
-{\rm d}\sigma^{VS,1}_{NNLO}\right)\nonumber \\&&
+ \int_{{\rm d}\Phi_{5}}
{\rm d}\sigma^{S}_{NNLO}
+\int_{{\rm d}\Phi_{4}}{\rm d}\sigma^{VS,1}_{NNLO} 
+ \int_{{\rm d}\Phi_{3}}{\rm d}\sigma^{V,2}_{NNLO}\;,
\end{eqnarray*}
where ${\rm d} \sigma^{S}_{NNLO}$ denotes the real radiation subtraction term 
coinciding with the five-parton tree level cross section 
 ${\rm d} \sigma^{R}_{NNLO}$ in all singular limits~\cite{doubleun}. 
Likewise, ${\rm d} \sigma^{VS,1}_{NNLO}$
is the one-loop virtual subtraction term 
coinciding with the one-loop four-parton cross section 
 ${\rm d} \sigma^{V,1}_{NNLO}$ in all singular limits~\cite{onelstr}. 
Finally, the two-loop correction 
to the three-parton cross section is denoted by ${\rm d}\sigma^{V,2}_{NNLO}$.
With these, each line in the above equation is individually 
infrared finite, and 
can be integrated numerically.

Systematic methods to derive and integrate subtraction terms 
were available in the literature only to NLO~\cite{nlosub,ant},
with extension to NNLO in special cases~\cite{cshiggs}.
In the context of 
this project, we fully developed an NNLO subtraction formalism~\cite{ourant}, 
based on the antenna subtraction method originally proposed at 
NLO~\cite{cullen,ant}. 
The basic idea of the antenna subtraction approach is to construct 
the subtraction terms  from antenna functions. 
Each antenna function encapsulates 
all singular limits due to the 
 emission of one or two unresolved partons between two colour-connected hard
partons.
This construction exploits the universal factorisation of 
phase space and squared matrix elements in all unresolved limits.
The individual antenna functions are obtained by normalising 
three-parton and four-parton tree-level matrix elements and 
three-parton one-loop matrix elements 
to the corresponding two-parton 
tree-level matrix elements. Three different types of 
antenna functions are required,
corresponding to the different pairs of hard partons 
forming the antenna: quark-antiquark, quark-gluon and gluon-gluon antenna 
functions. All these can be derived systematically from matrix 
elements~\cite{our2j} for physical processes. 

The factorisation of the final state phase space into antenna phase 
space and hard phase space requires a mapping of the antenna momenta 
onto reduced hard momenta. We use the mapping derived in~\cite{dak1} for 
the three-parton and four-parton antenna functions. To extract the infrared 
poles of the subtraction terms, the antenna functions must be integrated 
analytically over the appropriate antenna phase spaces, which is done by 
reduction~\cite{babis} to known 
phase space master integrals~\cite{ggh}. 

A detailed description of the 
calculation will be given elsewhere~\cite{our3j}.

\section{Results}
The resulting numerical programme, {\tt EERAD3}, yields the full kinematical 
information on a given multi-parton final state. It 
can thus be used to compute any 
infrared-safe observable in $e^+e^-$ annihilation
related to three-particle final states at 
${\cal O}(\alpha_s^3)$. As a first application, we derived results for the 
NNLO corrections to the thrust distribution~\cite{ourT}.

\begin{figure}[t]
\epsfig{file=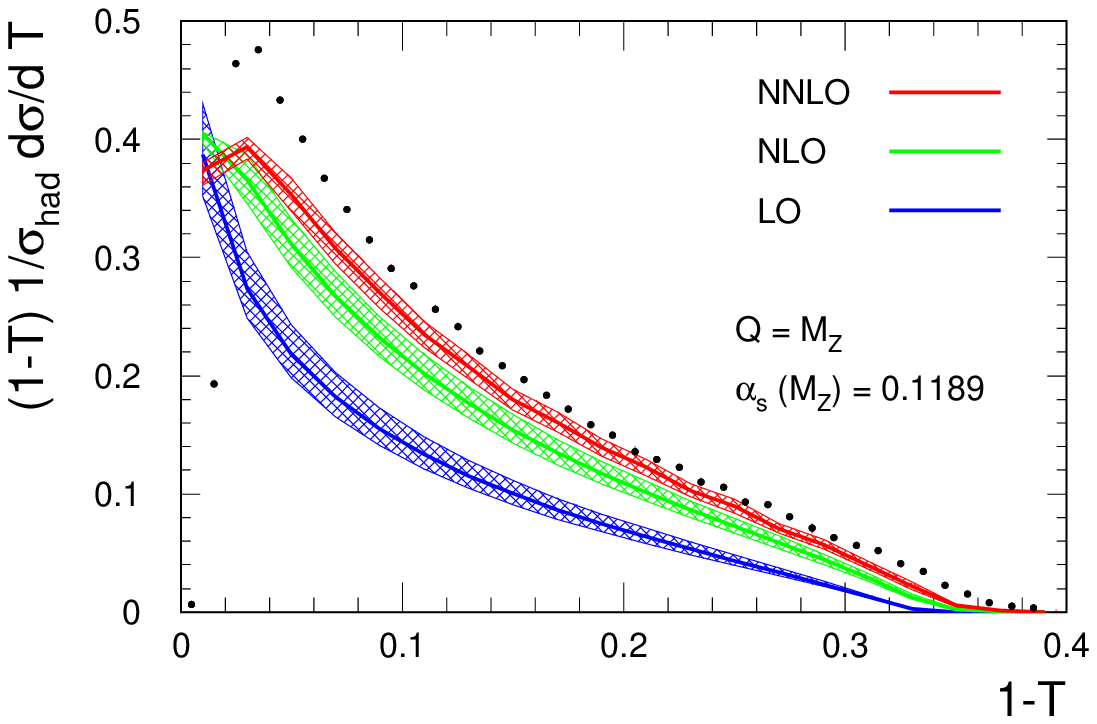,angle=-90,width=0.45\columnwidth}
\hspace{0.05\columnwidth}
\epsfig{file=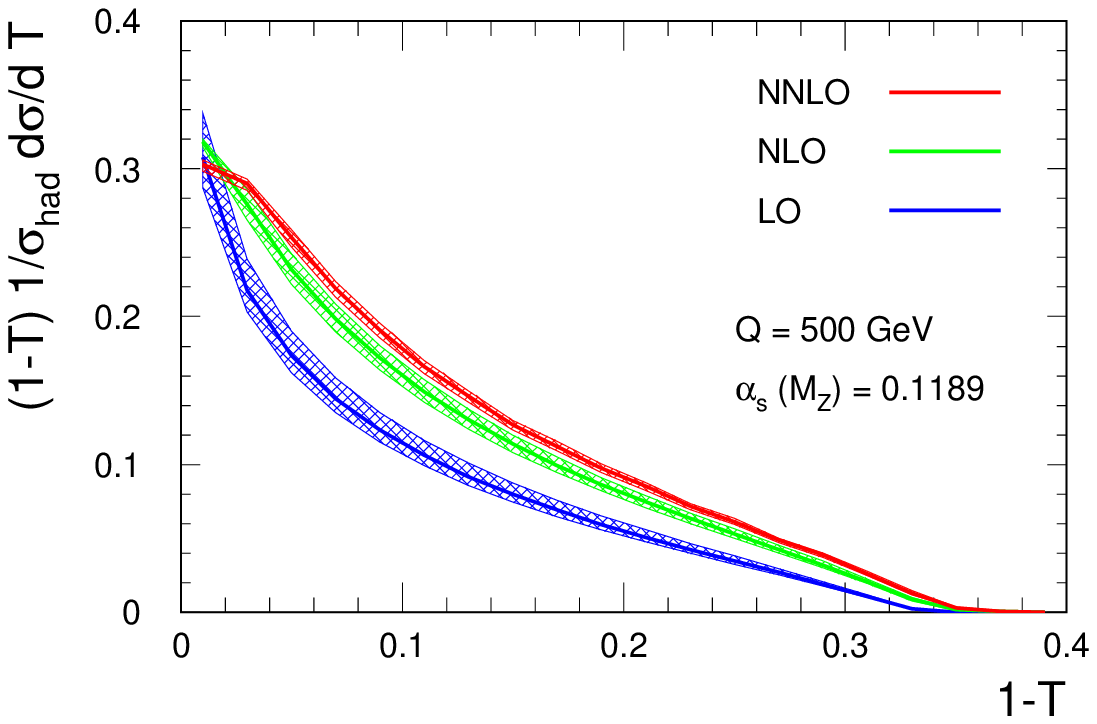,angle=-90,width=0.45\columnwidth}
\caption{Thrust distribution at LEP and at
the ILC with $Q= 500$~GeV.\label{fig:thrust}}
\end{figure}

In the numerical evaluation, we use $M_Z= 91.1876$~GeV and $\alpha_s(M_Z)=
0.1189$~\cite{Bethke}. Figure~\ref{fig:thrust} displays the perturbative 
expression for the thrust distribution at LO, NLO and NNLO, evaluated 
for LEP and ILC energies. The error band indicates the variation of the 
prediction under shifts of the renormalisation scale 
in the range $\mu \in [Q/2;2\,Q]$ around the $e^+e^-$ centre-of-mass 
energy $Q$.

It can be seen that even at linear collider energies,
inclusion of the NNLO corrections 
enhances the thrust distribution by around 10\% over the 
 range $0.03 < (1-T) < 0.33$, where relative scale uncertainty is
 reduced by about 30\% between NLO and NNLO. 
Outside this range, one does not expect the 
perturbative fixed-order prediction to yield reliable results. 
For $(1-T)\to 0$, 
the convergence of the perturbative series 
is spoilt by powers of
logarithms $\ln(1-T)$ appearing in higher perturbative orders, 
thus necessitating an all-order resummation of these logarithmic 
terms~\cite{ctwt,ctw}, and a matching of fixed-order and resummed 
predictions~\cite{hasko}.

The perturbative parton-level
prediction is compared with the hadron-level data from
the ALEPH collaboration~\cite{aleph}
 in Figure~\ref{fig:thrust}.  Similar data are also available from the
other LEP experiments~\cite{lep}. 
 The shape and
normalisation of the parton-level NNLO prediction agrees better with the
data than at NLO.
We also see that the NNLO corrections account
for approximately half of the difference between the parton-level NLO
prediction and the hadron-level data.  
 A full study including resummation of infrared logarithms
and
hadronisation corrections is underway.

\section{Conclusions}
We developed 
a numerical programme which can compute any infrared-safe observable
through to ${\cal O}(\alpha_s^3)$, which we applied here to determine the 
NNLO corrections to the thrust distribution. These corrections are moderate,
indicating the convergence of the perturbative expansion. Their inclusion 
results in a considerable reduction of the theoretical error on the 
thrust distribution. Our results will allow a 
significantly  improved determination of 
the strong coupling constant from jet observables from existing LEP data as 
well as from future ILC data.

\section*{Acknowledgements}
This research was supported in part by the Swiss National Science Foundation
(SNF) under contract 200020-109162, 
 by the UK Science and Technology Facilities Council and by the European 
Commission under contract MRTN-2006-035505 (Heptools).

\begin{footnotesize}

\end{footnotesize}
\end{document}